\documentclass[aps,pre,twocolumn,groupedaddress]{revtex4}
\usepackage{graphicx}
\def\be{\begin{equation}}
\def\ee{\end{equation}}
\begin{document}
\draft

\title{Image Effects on the Transport of Intense Nonaxisymmetric Charged Beams}
\author{Renato Pakter, 
Yan Levin and Felipe B.
Rizzato}
\address{Instituto de F\'{\i}sica,
Universidade Federal do Rio Grande do Sul\\
Caixa Postal 15051, 91501-970, Porto Alegre, RS, Brazil
}
\begin{abstract}
The effect of conducting pipes on the equilibrium of intense
nonaxisymmetric continuous beams of charged particles is investigated. 
For a cylindrical pipe and an elliptical beam, 
we obtain an exact closed form
analytical expression for the electrostatic potential. Using a variational 
principle, we then explore 
the distortions that equilibrium beams suffer due to the conducting channel. 
Finally, we present 
an exact proof
that despite the nonlinear forces acting on beams of 
arbitrary cross section inside
conducting pipes of arbitrary shape,
the density of these beams remains homogeneous and their cross sectional area remains the same as the one
in free-space.
\end{abstract}
%
\maketitle
%

In many applications where intense charged particle beams are employed, it is desirable
to have a nonaxisymmetric beam distribution. This is particularly the case for
vacuum electronic devices for which the use of ribbon electron beams allows for the transport
of a large amount of current at low space-charge forces \cite{stur59,zhang,boos93a,boos93b,chen06}. 
This is accomplished 
by distributing the current along
a {\it large} beam width, while keeping its height small enough to go across the small 
aperture sizes required by the modern high-frequency devices.
Although the focusing field configuration necessary to transport nonaxisymmetric beams
is naturally more complex than that of axisymmetric beams, a number of different schemes has been
proposed in the past few decades \cite{stur59,zhang,boos93a,boos93b}. They are all based on anisotropic 
periodic focusing fields which generate stronger focusing in one transverse direction than in another. 
Both stability studies \cite{boos93a,boos93b} and experiments with low-intensity beams \cite{zhang} show
the viability of the transport of such beams. More recently, Zhou et.al. \cite{chen06}
demonstrated the existence of a class of equilibrium solutions for the transport of 
intense, cold, nonaxisymmetric beams with variable aspect ratios through periodic magnetic focusing 
fields in free space. In equilibrium,
the beam is uniformly distributed along an ellipse whose angle and semi-axis radii undergo 
some small-amplitude rapid oscillations around stationary, average, values. 
A very important question then arises:
how will the equilibrium beam configuration be affected by the presence of
a conducting wall of a vacuum chamber? While round beams will not be disturbed by the coaxial
circular conducting pipe --- such pipe is an equipotential surface --- elliptical beams may
be strongly modified in such environment. In particular, one  might expect that the charge 
induced on a grounded conducting wall will strongly attract the beam particles, causing 
a significant modification of the beam equilibrium shape and its homogeneity.

In this paper, we study in detail the effects of a conducting pipe on the equilibrium of intense
nonaxisymmetric beams. We start by analyzing the image effects of a cylindrical 
conducting pipe on a continuous beam of an elliptical form. In contrast to previous
studies which employed multipole expansions \cite{reiser96,chen03}, we derive an explicit
analytical expression for the self-field potential of the beam inside the pipe. 
By means of a variational method we then calculate the relaxed equilibrium shape after a 
free-space beam enters into a conducting pipe. 
Finally, we prove that despite 
the nonlinear forces  produced by 
the conducting walls of {\it arbitrary cross section}, intense beams preserve their homogeneity
and conserve the cross sectional area.

We consider an intense, cold, unbunched beam propagating with an axial
velocity $v_z$ through a
magnetic focusing channel enclosed by a conducting pipe, both 
aligned with the $z$ axis. 
The focusing force is assumed to be linear
and anisotropic along the transverse directions.
In the smooth-beam 
approximation, where the fast oscillations due to the periodic
focusing field are averaged out, the dynamics of any beam particle is governed by \cite{boos93a,davidson}
\be
{\bf r}''+\nabla_\perp U_B+\nabla_\perp\psi=0,
\label{ri}
\ee
where ${\bf r}=x{\bf \hat e}_x+y{\bf \hat e}_y$,
$r=(x^2+y^2)^{1/2}$ is the radial distance from the $z$ axis,
the prime denotes derivative with respect to $z$,
$\nabla_\perp \equiv (\partial/\partial x){\bf \hat e}_x+(\partial/\partial y){\bf \hat e}_y$,
$U_B=k_x x^2/2+k_y y^2/2$ is the effective confining potential due to the external field,
$k_i=\xi _i^2q^2\overline{B(z)^2}/2\gamma _b^2\beta_b^2m^2c^4$, $i=x,y$,
$B(z)$ is the magnetic field along the $z$ axis, the bar represents average over one focusing period, $\xi _i$ are the form
factors which satisfy $\xi _x+\xi _y=1$, $\beta _b=v_z/c$, $\gamma_b=(1-\beta _b^2)^{-1/2}$, $q$ and $m$ are the mass and charge of 
the beam particles, and $c$ is the speed of light in {\it vacuo}. 
In Eq. (\ref{ri}), $\psi$ is a normalized potential that incorporates both
the self-electric and the self-magnetic fields and is also affected by the presence of a conducting 
wall. It is related
to the self-scalar and self-vector potentials by
$\phi^s=\beta_b^{-1}A_z^s=\gamma_b^3m\beta_b^2c^2\psi({\bf r},s)/q$ and satisfies the Poisson Equation
\be
\nabla_\perp ^2\psi=-{2 \pi K\over N_b} n_b({\bf r},z),
\label{poi}
\ee
subjected to the boundary condition $\psi=0$ at the grounded conducting wall. 
Here, $n_b({\bf r},z)$ is the beam density profile, $N_b=$ const. is the number of
particles per unit axial length, and $K=2q^2N_b/\gamma_b^3\beta_b^2mc^2$ is
the so-called beam perveance which can be interpreted as a measure of the total 
two-dimensional beam charge.

We begin by studying how to include the effects of cylindrical conducting pipe of radius $r_w$ in the
self-field potential. Let us
consider an arbitrary transverse particle distribution of total charge $K$
contained in the region $r<r_w$. In the free-space, the electrostatic potential satisfies the Laplace equation 
$\nabla_\perp ^2\psi^{free}=0$ for $r>r_w$. We assume that $\psi^{free}$ is known. One can then verify that the function
$\psi^{free}(r_w^2/r,\theta)$ also satisfies the Laplace equation.  This function, however, has a singular point of
charge $K$ at $r=0$, as well as other singularities --- corresponding to the image charges --- located at $r>r_w$.  
We then note that the combination
\be
\psi({\bf r})=\psi^{free}(r,\theta)-\psi^{free}(r_w^2/r,\theta)-K \log (r/r_w),
\label{psi}
\ee
satisfies the Poisson equation with the original charge 
density and vanishes at $r=r_w$. This means that $r_w$ corresponds to the location of a grounded 
conducting wall. Note that although we are
only interested in the electrostatic potential $\psi({\bf r})$ inside the pipe, to obtain it, it is necessary
to know $\psi^{free}({\bf r})$ over {\it the whole} space.

We employ the above result to investigate the transport of an elliptically symmetric inhomogeneous beam
propagating inside a conducting pipe of radius $r_w$. We assume a parabolic density
profile of the form
\be
n(x,y)={N_b\over \pi a b}\left [1+\chi-2\chi\left({x^2\over a^2}+{y^2\over b^2}\right )\right]
\ee
inside the beam core, $(x/a)^2+(y/b)^2\leq 1$, where $\chi$ is the inhomogeneity parameter ($-1\le\chi\le 1$), and
$a$ and $b$ are the elliptical semi-axis radii. In the absence of a conducting wall, a free space analytic solution 
to the Poisson Equation, $\psi^{free}({\bf r})$, is known inside the beam core \cite{davidson,kosc}.
Solving the Laplace equation in elliptical coordinates, imposing the continuity of the electric field 
at the beam boundary, and taking advantage of complex variable properties, we can also obtain a closed-form expression
for the self-field outside the beam core
\be
\psi^{free}({\bf r})=Re\left[{3+\chi\over 6}
{\cal H}+{\chi\over 12}{\cal H}^2
-{\rm Arccosh}\left({\zeta\over c}\right)\right],
\label{psiout}
\ee
where $\zeta=x+iy$, ${\cal H}= 1-2(\zeta/c)^2[1-\sqrt{1-(c/\zeta)^2}]$, and $c= \sqrt{a^2-b^2}$.
In the limit of $c\rightarrow 0$, $\psi^{free}\rightarrow \log (2r/c)$, as expected.
Substituting $\psi^{free}({\bf r})$ in Eq. (\ref{psi}), we find an exact analytic expression
for the electrostatic potential of an inhomogeneous elliptic beam inside a grounded conducting pipe.
The electric field generated by such a confined beam can be then calculated
and used to study, for instance, the dynamics of test particles \cite{ikegami99}. 
Here we will use the analytical expressions and a variational calculation to 
investigate the effects of conductors on  
the equilibrium beam profile.

\begin{figure} [ht]
\includegraphics[scale=1]{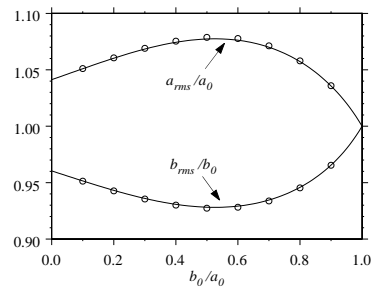}
\caption{Equilibrium beam dimensions as a function of the aspect ratio $b_0/a_0$ for $r_w/a_0=1.2$.
The curves correspond to the results of the variational calculation, whereas the symbols,
to those obtained from the full N-particle simulations.}
\end{figure}

For a given set of transport channel parameters
$k_x$, $k_y$, $r_w$  
we compute the total beam energy per particle
\be
{\cal E}_T={1\over N_b}\int \left[{\psi({\bf r})\over 2}+U_B({\bf r})\>\right] n({\bf r})\> d^2r,
\ee
as a function of the beam parameters $a$, $b$ and $\chi$. 
By minimizing ${\cal E}_T(a,b,\chi)$ with respect to
these parameters, 
we then determine the equilibrium beam shape inside the pipe. Note that in this variational calculation, 
the beam shape is constrained to be elliptical, 
while its inhomogeneity is allowed to vary parabolically.  
For $r_w\rightarrow\infty$ the minimization can be performed explicitly to find that the 
equilibrium corresponds to a uniform beam ($\chi=0$)
with $a=a_0\equiv \sqrt{2Kk_y/[k_x (k_x+k_y)]}$ and $b=b_0\equiv \sqrt{2Kk_x/[k_y (k_x+k_y)]}$. 
The free space radii $a_0$ and $b_0$ can then be used to characterize the focusing field
intensity $k_x$ and $k_y$.
In Fig. 1 we show the
results obtained for $r_w/a_0=1.2$ and varying values of $b_0/a_0$. For later comparison with the full N-particle simulations,
we present in panel (a) the equilibrium {\it effective} semi-axis $a_{rms}\equiv 2\langle x^2\rangle ^{1/2}=a(1-\chi/3)^{1/2}$ and 
$b_{rms}\equiv 2\langle y^2\rangle ^{1/2}=b(1-\chi/3)^{1/2}$, where $\langle \cdots\rangle$ stands for the average over the beam distribution.
The figure confirms that for nearly axisymmetric beams with $b_0/a_0 \approx 1$, wall effects are negligible and $a_{rms}/a_0$
and $b_{rms}/b_0$ are close to unity. As the focusing channel becomes more anisotropic with $b_0/a_0<1$, 
wall effects become important, always acting to further intensify the beam anisotropy. 
The figure also reveals that the dependence of the equilibrium beam sizes on the focusing field anisotropy $b_0/a_0$
is non monotonic, being more pronounced for aspect ratios close to $b_0/a_0=0.5$. This 
feature was verified for different wall positions $r_w$. In all the cases, the 
inhomogeneity parameter $\chi$ was found to be small, on the order of a few percent.
Full N-particle simulations were
performed to compare with the results of the variational calculations. In the simulations, a large number $N=20000$ of 
macroparticles evolved according to Eq. (\ref{ri}). The influence of a grounded conducting 
wall was taken into account using the image
charges \cite{karen}.  The particles were launched in an arbitrary 
configuration and attained equilibrium state through a slow damping in their dynamics. The simulation results obtained for 
$a_{rms}$ and $b_{rms}$ are represented by the symbols in Fig. 1(a), showing a very good agreement with the predictions of the
variational calculation. 
An intriguing property of the variational calculation 
is that the effective area occupied by
the beam in the presence of a wall is {\it exactly} the same as that of the free beam, i.e., $a_{rms}b_{rms}=a_0 b_0$ 
[see Fig. 1(a)]. This feature in a variational calculation
suggests a hidden symmetry which, however, can not be seen at the level of
variational equations because of their complexity.

To uncover the hidden symmetry we now appeal to the following mathematical adiabatic construction.
Consider an equilibrium beam in the absence of a wall. Because each particle is in equilibrium, the force balance
Eq. (\ref{ri}) requires that
 \be
\nabla_\perp U_B+\nabla_\perp\psi=0
\label{equi2a}
\ee 
If we picture the beam as a continuous
charge distribution, this equation holds at {\it all} points inside the
the beam distribution. 
We now suppose that a conducting boundary of an arbitrary shape
initially at infinity is adiabatically approaching the beam (shrinking towards it). 
The equilibrium beam charge distribution deforms, but 
since the wall motion is adiabatic, the equilibrium condition Eq.(\ref{equi2a}) remains unaltered. 
Operating with $\nabla_\perp$ on Eq.(\ref{equi2a}) we obtain
\be
\nabla_\perp^2U_B({\bf r})-{2 \pi K\over N_b} n_b({\bf r},r_w)=0,
\label{equi1}
\ee
where $r_w$ now stands for a typical distance from the focusing channel axis to the conductor and 
use has been made of the Poisson equation (\ref{poi}). Note that while
$n_b$ is a function of $r_w$,  $U_B({\bf r})$ only depends
on the external focusing field. As the wall undergoes a small displacement $\delta r_w$ 
with $r_w\rightarrow r_w+\delta r_w$, the beam particle positions will be 
modified, ${\bf r}\rightarrow {\bf r}+\delta {\bf r}$.
Using this in Eq.~(\ref{equi1}) and expanding to 
linear order in $\delta r_w$ and $\delta {\bf r}$ we obtain
\be
\left[{\partial\over\partial r_w}+{\bf v}.\nabla_\perp\right ]n_b({\bf r},r_w)=
{N_b\over 2 \pi K}{\bf v}.\nabla_\perp\left(\nabla_\perp^2U_B\right ),
\label{equi2}
\ee
where ${\bf v}\equiv \delta{\bf r}/\delta r_w$. 
Thus, as long as $U_B$ is a quadratic function of ${\bf r}$ --  the focusing force is linear --
the right-hand-side of Eq. (\ref{equi1}) vanishes and the total (convective) derivative of $n_b({\bf r})$ 
with respect to variations in $r_w$ is zero. Since in the absence of the wall the density is uniform, vanishing of the convective
derivative implies that it remains so for any $r_w$, preserving the beam cross sectional area. Therefore,
the beam density and cross-sectional area are adiabatic invariants.
This places a stringent constraint on the
effects that conductors can exert on beam equilibria. To illustrate this,
in Fig. 3 we show the large distortion of the equilibrium charge distribution caused by a nearby
conducting plate. Nevertheless, in agreement with Eq. (\ref{equi2}) the particle density inside the beam
remains uniform and constant, and the cross sectional area of the beam is unchanged.

\begin{figure} [ht]
\includegraphics[width=7cm]{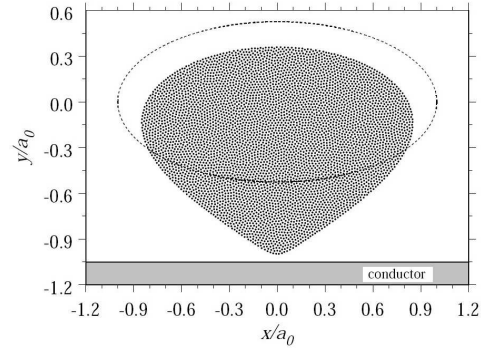}
\caption{Equilibrium distribution in a focusing field with $b_0/a_0=1/1.9$ near a conducting plate
obtained using a full N-particle simulation. The dashed ellipse shows the equilibrium beam border in the
absence of the conducting plate.}
\end{figure}

To conclude, we have investigated the effects of a conducting pipe on the equilibrium of intense
nonaxisymmetric beams. First, we analyzed the image effects of a cylindrical 
conducting pipe on a continuous beam with elliptical symmetry and derived an exact 
expression for the electrostatic potential. 
Using a variational method, we then calculated the equilibrium beam shape and its charge
distribution. It was found that
the presence 
of a pipe does not alter the effective beam cross sectional area. 
This suggested that the variational equations possess an underlying hidden symmetry. 
Using an adiabatic construction we were able to prove that despite 
the nonlinear forces exerted by the induced charges, 
the intense particle beams preserve a uniform equilibrium  
density, as long as the focusing forces are linear. Furthermore, the
cross sectional area of the beam remains the same as in the absence of a conductor. 
These findings  
should have important practical implications for the
design of intense beam transport channels.

Work supported by CNPq and FAPERGS, Brazil, and by the US-AFOSR, 
grant number FA9550-06-1-0345. 




\end{document}